# Low-Cost Turntable Designed for RF Phased Array Antenna Active Element Pattern Measurement

Rebekah Edwards, Taylor Martini, Jonathan E. Swindell, David W. Cox, Adam C. Goad,
Austin Egbert, Charles Baylis, Robert J. Marks
Department of Electrical and Computer Engineering, Baylor University, USA
Charles_Baylis@baylor.edu

*Abstract*—Accurate antenna array calibrations and measurements of aspects such as active element pattern (AEP) are critical for enabling integrated sensing and communication (ISAC) technologies such as directional modulation. One reliable way of obtaining accurate and repeatable AEP measurements is to spin the antenna array on a turntable, but many turntables designed for antenna array measurements are prohibitively expensive for small labs and may not be designed with RF considerations, such as cable phase stability, in mind. This paper details the design of a motorized 3D printed turntable for use in directional modulation and in-situ measurement experiments that will allow for rotation of an antenna array around a point, such that the far field of the antenna pattern can be measured by a stationary receiver.

## I. INTRODUCTION

Active element pattern (AEP) is widely used in antenna array theory and applications. Using AEP, the total array pattern is the superposition of the contributions of exciting each antenna with a given input in the presence of the array, given that the remaining elements are terminated in a matched load [1, 2]. Constructing the array pattern from AEP measurements is advantageous because it accounts for mutual coupling effects caused by other antennas in the array reradiating electromagnetic energy transmitted from a given antenna, while still taking advantage of superposition to reduce the number of measurements required to construct the array pattern. AEP is used widely in array calibration and applications. Pozar presents a method for obtaining the scattering parameters of all elements of a phased array from the array's active element pattern [3]. Daly and Bernhard use directional modulation to enable the transmission of multiple communication or radar waveforms to be transmitted in multiple directions simultaneously using a single phased array antenna, using AEP to account for mutual coupling effects [4]. Daly and Bernhard have extended directional modulation to beam steering applications [5].

To leverage these applications, AEP must be very accurate for a given array. Collecting high-fidelity AEP measurements requires expensive equipment and time-consuming measurements. Salas-Natera provides an excellent overview of the measurement, characterization, and calibration of active antenna arrays, which highlights the cost of measurement acquisition and that automating measurements can save costs [6]. Wei demonstrates a fast measurement method that reduces the number of measurements required to determine an active element pattern [7].

Simulations can decrease the time and resources needed to acquire AEP, but at the cost of model accuracy. When measuring AEP in a well-controlled environment, the measurement equipment uncertainty limits the accuracy of AEP. Electromagnetic simulation software approximates solutions to Maxwell's Equations, which can yield accurate simulation results but take a long time to run. Research has been done on surrogate models that estimate AEP faster than HFSS. While these methods can be useful for a given application, it is important to remember that they make additional assumptions about AEP that can increase error. Zhang presents a method to calculate scattering parameters for a large finite array given an active element pattern that reduces the computational cost and shows good agreement with HFSS simulations [8]. Yang found that using active element patterns can reduce the computational time required to simulate the radiation pattern compared to full-wave methods, especially when the array is very large [9]. Huang and Liu demonstrate a method that simulates the active element pattern of a linearly spaced array and then extends that to support unequally spaced arrays [10, 11]. Yang applies the K-nearest neighbor algorithm to predict the active element pattern of an array with arbitrary array geometries and a varying number of elements [12]. Hong applies a knowledge-based neural network to act as a surrogate model to predict AEP based on the distance between elements and antenna properties [13].

Despite advances in AEP simulation methods, measurement remains the gold standard. Turntables are commonly used in antenna measurements, including AEP, because they provide precise and repeatable angular positioning to antennas under test (AUT) [14, 15]. Even for turntables with small beds of around 0.3 meters, commercial-off-the-shelf (COTS) turntables are often tens of thousands of dollars [16, 17, 18]. Another disadvantage of these turntables is that they are not designed for antenna arrays and may require additional assembly on top of the turntable to support measuring an array. The contribution of this paper is a low-cost turntable designed to automate AEP measurement collection and prevent electrical issues that could decrease the fidelity of AEP measurements.

## II. TURNTABLE DESIGN

### A. Design Requirements

The goal of this endeavor was to design a motorized 3D printed turntable for use in directional modulation and in-situ measurement experiments that will allow for rotation of an antenna array around a point, such that the far field of the antenna pattern can be measured by a stationary receiver. Based on the intended use and desired features of the turntable, in addition to the challenges faced with the development of the design, the following requirements for the turntable design were established:

1. The turntable top shall be able to freely and smoothly rotate at least 180 degrees without obstruction or excessive friction.
2. The center point of the antenna array shall be the center point of the turntable.
3. The turntable top shall support the weight of two SigaTek dual-directional couplers (20dB; 2.0-4.0 GHz; Part Number SCC2012142), two monopole antennas, and the cables and connectors needed for operation.
4. The turntable base shall support the weight of the turntable top and the motor used to turn it.
5. The turntable base shall be resistant to heat generated by the motor.
6. The couplers and antennas shall be securely fixed relative to the turntable top.
7. The turntable shall have the ability to be affixed to the top of a tripod.

In addition to the requirements, the goals for the design of the system were as follows:

1. Minimize cable loading on the couplers to preserve their structural integrity.
2. Minimize cable-induced torque and mechanical loading transmitted to the motor to reduce required motor torque.
3. Prioritize ease of assembly.
4. Provide a mechanism to level the turntable.
5. Provide a mechanism of visually indicating the angular position of the turntable.
6. Ensure phase and calibration stability of measurements.
7. Minimize electromagnetic scattering and interference.
8. Ensure system adaptability to accommodate alternative antenna arrays.

### B. Design Process

There were three relevant iterations of the original turntable. The original 3D printed design was a motorized turntable that held two antennas and two DigiKey FPC07180 directional couplers [19] in place and was mounted to a tripod. While this design worked to rotate the system, several issues were encountered. These issues included the breakage of the antenna holder due to a split along the 3D printed layer lines, the exposure of the couplers to significant amounts of torque due to the cables extending directly out from the sides of the turntable, and the inability of the motor mount to hold the turntable steady and prevent tilting. These couplers were constructed on a thin layer of quartz, which made them extremely fragile. The quartz layers often fractured or even shattered if any twisting force from a cable was put on the coupler substrate.

The first iteration developed a sturdier motor mount, stronger antenna holders, torque-relieving boxes to hold the couplers, and added right-angle SMA connectors after the couplers to allow the cables to point downward and lessen the torque caused by the cables extending directly out from the sides. However, in order to rotate the system, the motor still had to overcome the resistive torque applied by the cables, causing it to overheat and warp the PLA motor mount. This caused the system to tilt and become unstable.

In the second iteration, the motor mount was 3D printed out of resin instead of PLA. This solved the overheating problem but did not reduce any cable-induced loading applied to the motor or couplers, causing the couplers to break during use. Because the main structural design of the turntable remained the same, the couplers still shattered due to the torque caused by the cables. Additionally, testing revealed that the right-angle SMA connectors caused reflections that changed often and could not be calibrated out. A full overhaul of the design was needed.

### C. Final Design

The third iteration of the turntable is what is presented in this paper. The model was modified to use SigaTek dual-directional couplers, which are larger and more mechanically robust than their predecessors. A two-level system was implemented to remove the structural load from the motor, instead of the original single-level design where the motor directly supports the rotating assembly. In this design, the motor is now mounted to the turntable base, while the rotating component, the turntable top, is supported by roller bearings to carry the structural weight of the system. This design reduces the required torque from the motor, improves rotational smoothness, and limits the transmission of cable-induced loading to the motor and couplers. A new modular tripod mount was developed to allow the turntable assembly to be adapted to a different tripod or set on a flat surface if needed. The locations of the hanging cables, previously the largest issue faced due to the torque they created, were moved as close as possible to the center of the turntable (with clearance around the motor) to allow the turntable to spin a full 180 degrees unobstructed. Finally, angle markers were added to the turntable top to make the visual assessment of the rotation easier, and a mini circular level was added to ensure proper installation. The turntable is comprised of six 3D printed components; five were made of PLA and printed with an UltiMaker S5 with a 330x240 mm print bed, while the motor mount was printed with resin.

Fig. 1 shows an isometric view of the 3D modeled turntable assembly. The small, yellow circular level can be seen near the dovetail connection of the turntable top, and angle markers can be seen ringing the edges of the turntable. Fig. 2 shows an isometric view of the turntable assembly with a transparent top, allowing the lower-level components to be seen. The bearings support the weight of the turntable top and can be seen attached at the top of each leg of the turntable base. Fig. 3 shows a right-side view of the turntable assembly. The blue cables are semi-rigid and do not move, whereas the red cables are flexible, phase-stable cables, and will move slightly as the turntable rotates. The tripod mount is the lowest component in this view. Fig. 4 shows an isometric bottom view of the turntable assembly. The tripod mount can be clearly seen, as can the red phase-stable cables. These cables extend out of the base of the turntable and are plugged into the RFSoC.

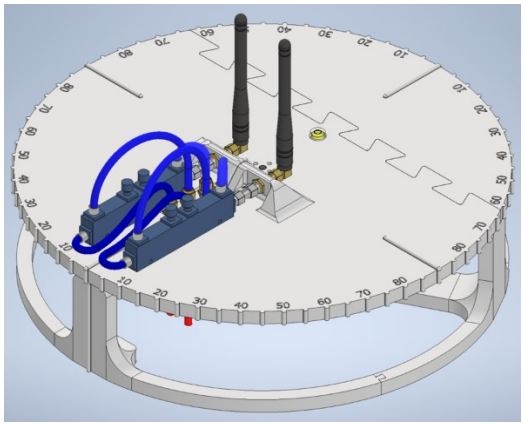

Fig. 1. An isometric view of the turntable assembly.

Fig. 5 shows the turntable top assembly. Due to the 3D printing bed constraints of 330x240 mm, the 324 mm turntable was split into two components that are connected using the dovetail joining method. Fig. 6 shows the turntable base assembly. Like the turntable top assembly, the base assembly was split into two, front and back, due to the 3D printing build plate constraints. The turntable base was designed with the dovetail joining method, placements to secure the resin motor mount to the top side and the tripod mount to the underside, as well as bearing mounts on the three vertical supports.

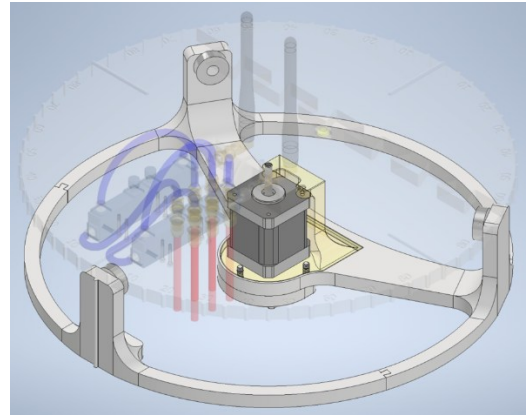

Fig. 2. An isometric view of the turntable assembly with transparent top so that the lower components can be clearly seen.

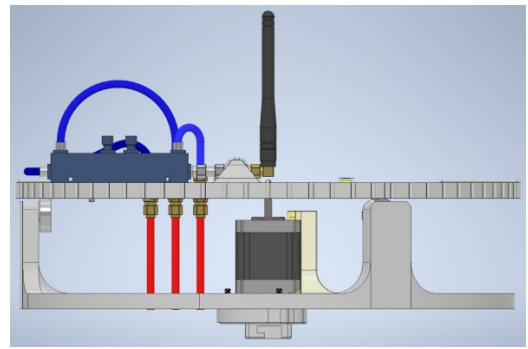

Fig. 3. A right-side view of the turntable assembly.

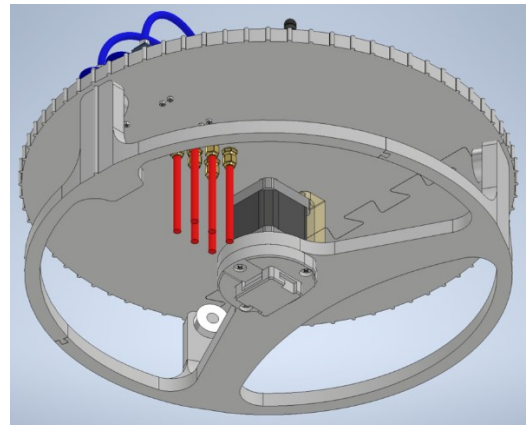

Fig. 4. An isometric bottom view of the turntable assembly.

This turntable design is powered by a NEMA 17 stepper motor with 68 oz·in of torque [25] and is driven by a SparkFun ProDriver TC78H670FTG Stepper Motor Driver [26]. One full step rotates the motor by 0.9 degrees, but greater rotation resolution can be achieved using microstepping. This microstepping can achieve a precision of 0.00703125 degrees. The turntable can be automated using MATLAB via an Arduino board.

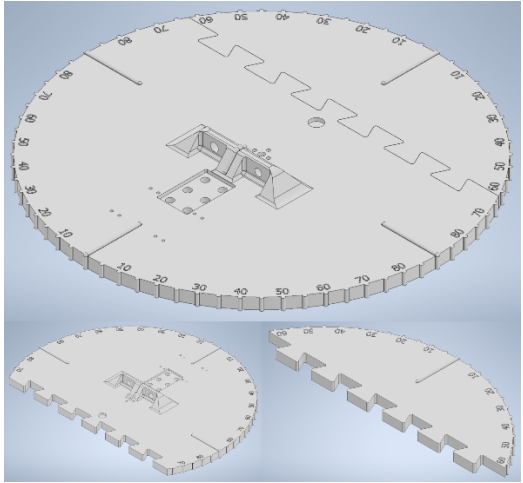

Fig. 5. An isometric view of the turntable top assembly, the turntable top back, and the turntable top front.

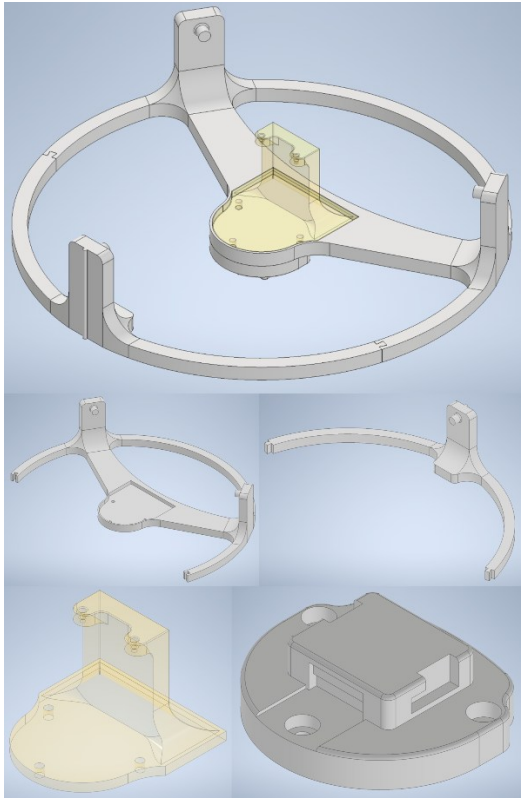

Fig. 6. An isometric view of the turntable base assembly, the turntable base front, the turntable base back, the turntable base motor mount (colored translucent yellow to indicate that it will be printed with resin), and the tripod mount that will be fastened to the bottom of the turntable base.

## III. RESULTS AND DISCUSSIONS

Fig. 7 shows photographs of the fully constructed turntable. The assembly is complete with the two couplers, the semi-rigid cables, and the antennas.

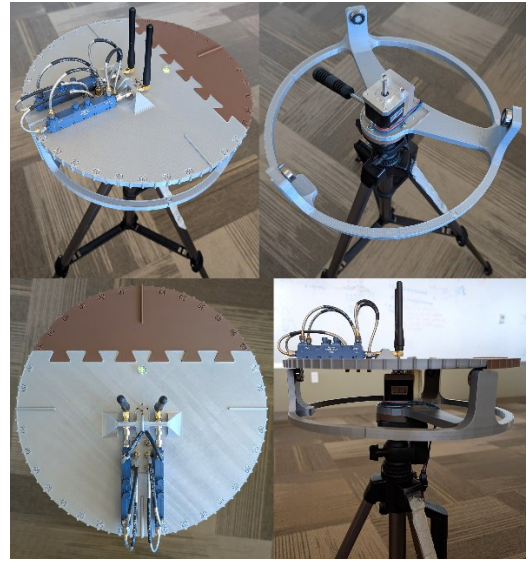

Fig. 7. Images of the constructed turntable: a perspective view, a perspective view with top removed to show the turntable base, a top view, and a right-side view.

When attempting to take AEP measurements, one significant source of error can be the lack of phase stability and calibration stability. While phase-stable cables can significantly help alleviate this issue, these cables are also heavy and have low flexibility. This force was typically caused by the large phase-stable cables that were connected to the four ports of the coupler. The presented turntable was therefore constructed with different couplers that were larger and much more robust, in addition to small rigid cables that connected the couplers to SMA male-to-male throughs. These throughs were mounted into the turntable, and the phase-stable cables were attached to the other side of them, ensuring that the force is applied only upon the turntable and not on the couplers. Additionally, the location of the cables has been placed as close as possible to the center of the axis of rotation of the turntable in order to minimize the resistive torque on the turntable.

Another important AEP consideration is the need to keep the antennas away from objects that might cause electromagnetic scattering and interference. Mounting the turntable on a tripod helps this significantly by providing both distance from the floor and the ability to move the setup to a better location, but the turntable must be kept level and stable in order to obtain accurate measurements. The use of a strong and sturdy tripod, combined with the implementation of a turntable base to stabilize the rotation of the turntable top, greatly improved the stability of the turntable. Table I compares the proposed turntable to three commercial off-the-shelf (COTS) turntables. Perhaps the most outstanding feature of this turntable-based design is that it is much less expensive than the others, providing excellent rotation resolution. While the rotation angle is only 180 degrees, this still provides significant ability to assess directional transmissions. The weight capacity was not tested to failure for this design; the listed 1.4 kg amount is the maximum load under which it was tested. Its tested weight capacity is comparable to one of the listed COTS setups. Although it is less than some of the other solutions, it is still sufficient to hold the transmitter array setup that is planned for the experiments. Table II shows the Bill of Materials (BoM) for the proposed turntable. Table III shows the RF and electronic components used for the turntable to make AEP measurements; these are not shown in the BoM as they are required for any turntable setup.

TABLE I
COTS TURNTABLE COMPARISON

| Manufacturer | Diameter | Price | Weight Capacity | Rotation Resolution | Rotation Angle |
|---|---|---|---|---|---|
| [16] Maturo | 0.3 m | $8,969 | 50 kg | 0.05° | 600° |
| [17] mmWave Test Solutions | 0.26 m | €7,800 ≈$9,377 | 1.2 kg | 0.01° | 360° (∞) |
| [18] Diamond Engineering | 1 ft ≈0.305 m | $15,000 | 10 lbs. ≈4.5 kg | 0.125° | 360° (∞) |
| Proposed Turntable | 0.31 m | $112 | 3 lbs. ≈ 1.4 kg | 0.007°** | 180° |

*Note: Prices were obtained via quotes on Feb. 11 and 12, 2026, and were converted to USD at the exchange rate on Feb. 12, 2026.
**Using micro-stepping with the motor; open-loop control.

TABLE II
PROPOSED TURNTABLE BILL OF MATERIALS

| Manufacturer | Component | Unit Price | Quantity | Total Price |
|---|---|---|---|---|
| [21] UltiMaker | PLA Filament | $0.052 | 569 g | $29.59 |
| [22] DigiKey | Through Connectors | $4.50 | 8 | $36.00 |
| [23] SHKI | 608 Bearings | $0.35 | 3 | $1.05 |
| [24] JYK | Mini Circular Level | $0.35 | 1 | $0.35 |
| [25] WanTai | Stepper Motor | $16.24 | 1 | $16.24 |
| [26] SparkFun | Motor Controller | $22.95 | 1 | $22.95 |
| [27] Pololu | Motor Shaft Collar | $5.75 | 1 | $5.75 |
| | | | Total Turntable Cost | $111.93 |

*Note: Prices are from the purchase date and may have changed.

TABLE III
RF AND ELECTRONIC COMPONENTS

| Manufacturer | Component | Quantity |
|---|---|---|
| [20] Sigatek | Dual-Directional Couplers | 2 |
| [28] Maury Microwave | Phase-Stable Cables | 6 |
| [29] Florida RF | Semi-Rigid Cables | 6 |
| * | Monopole Antennas | 2 |
| * | M-M SMA Adapters | 2 |

*Standard components made by many manufacturers.

While this design is sufficient for the current needs, future designs should consider the following:
1. Motor efficiency and the heat generated from operating the system. Alternative motors could be more efficient and generate less heat in the process should be investigated as well as a more in-depth look at the materials in contact with the motor. Additionally, heat sinks could be implemented on future designs to help disperse the heat.
2. Additional research should be done on manufacturing processes with the potential of using laser cut material or a CNC machine to fashion components instead of relying solely on 3D printing. Likewise, when 3D printing, the size compatibility between the 3D printer and turntable design needs to be ensured.

IV. CONCLUSIONS

A low-cost 3D printable turntable design specialized for antenna array measurements has been demonstrated. The design provides significantly lower costs than other published and available options. This design provides an excellent option for directional measurements to be performed at low cost. Additional issues that are expected to be addressed in future design iterations include enhanced motor efficiency and manufacturing process resolution.

ACKNOWLEDGMENT

The author wishes to acknowledge Adam Swinney and Ava Lea from Dr. Trevor Fleck's Additive Manufacturing Lab in the Department of Mechanical Engineering at Baylor University for their assistance in 3D printing the turntable components during all iterations of the project.